\documentclass[a4paper,twocolumn]{article}
\usepackage{amsfonts}

\begin{document}

\title{{\LARGE A MATHEMATICAL\ FOUNDATION\ OF\ QUANTUM\ INFORMATION AND QUANTUM\
COMPUTER}\\
{\LARGE -on quantum mutual entropy and entanglement-}}
\author{Masanori Ohya \\
Department of Information Sciences\\
Science University of Tokyo\\
Noda city, Chiba 278-8510, Japan}
\date{}
\maketitle

\pagestyle{empty}

\section{Introduction}

The study of mutual entropy (information) and capacity in classical system
was extensively done after Shannon by several authors like Kolmogorov \cite
{Kol} and Gelfand \cite{GY}. In quantum systems, there have been several
definitions of the mutual entropy for classical input and quantum output
\cite{BS,Hol,Ing,Lev}. In 1983, the author defined \cite{O2} the fully
quantum mechanical mutual entropy by means of the relative entropy of Umegaki%
\cite{U}, and it has been used to compute the capacity of quantum channel
for quantum communication process; quantum input-quantum output \cite{OPW1}.

Recently, a correlated state in quantum syatems, so-called quantum entangled
state or quantum entanglement, are used to study quntum information, in
particular, quantum computation, quantum teleportation, quantum cryptography
\cite{BO,Ben,BBPSSW,Eke,IOS,JB,Sch1,Sch2}.

In this paper, we mainly discuss three things below:(1) We point out the
difference between the capacity of quantum channel and that of
classical-quantum-classical channel followed from \cite{O11}. (2) So far the
entangled state is merely defined as a non-separable state, we give a wider
definition of the entangled state and classify the entangled states into
three categories. (3)The quantum mutual entropy for an entangled state is
discussed. The above (2) and (3) are a joint work with Belavkin\cite{BO} .

\vspace{-4mm}

\section{Qunatum Mutual Entropy}

The quantum mutual entropy was introduced in \cite{O2} for a quantum input
and quantum output, namely, for a purely quantum channel, and it was
generalized for a general quantum system described by C*-algebraic
terminology\cite{O4}. We here review the quantum mutual entropy in usual
quantum system described by a Hilbert space.

Let $\mathcal{H}$ be a Hilbert space for an input space, $B%
\mathcal{(H)}$ be the set of all bounded linear operators on $\mathcal{H}$
and $\mathcal{\ S(H)}${\scriptsize \ }be the set of all density operators on
$\mathcal{H}.$ An output space is described by another Hilbert space $%
\stackrel{\sim }{\mathcal{H}}$ , but often $\mathcal{H=}\stackrel{\sim }{%
\mathcal{H}}$. A channel from the input system to the output system is a
mapping $\Lambda $* from $\mathcal{S(H)}$ to $\mathcal{S(\stackrel{\sim }{%
\mathcal{H}})}$ \cite{O1}. A channel $\Lambda $* is said to be completely
positive if the dual map $\Lambda $ satisfies the following condition: $%
\Sigma _{k,j=1}^{n}$ $A_{k}^{*}\Lambda (B_{k}^{*}B_{j})A_{j}\geq 0$ for any
$ n\in {\bf N}$ and any $A_{j}\in B(\mathcal{H}),B_{j}\in
B(\stackrel{\sim  }{\mathcal{H}})$.

An input state $\rho $ $\in \mathcal{S(H)}$ is sent to the output system
through a channel $\Lambda $*, so that the output state is written as $%
\stackrel{\sim }{\rho }\equiv \Lambda ^{*}\rho .$ Then it is important to
ask how much information of $\rho $ is correctly sent to the output state $%
\Lambda ^{*}\rho .$ This amount of information transmitted from input to
output is expressed by the mutual entropy in Shannon's theory.

In order to define the quantum mutual entropy, we first mention the entropy
of a quantum state introduced by von Neumann. For a state $\rho ,$ there
exists a unique spectral decomposition $\rho =\Sigma _{k}\lambda _{k}P_{k},$%
where $\lambda _{k}$ is an eigenvalue of $\rho $ and $P_{k}$ is the
associated projection for each $\lambda _{k}$. The projection $P_{k}$ is not
one-dimensional when $\lambda _{k}$ is degenerated, so that the spectral
decomposition can be further decomposed into one-dimensional projections.
Such a decomposition is called a Schatten decomposition, namely, $\rho
=\Sigma _{k}\lambda _{k}E_{k},$where $E_{k}$ is the one-dimensional
projection associated with $\lambda _{k}$ and the degenerated eigenvalue $%
\lambda _{k}$ repeats dim$P_{k}$ times; for instance, if the eigenvalue $%
\lambda _{1}$has the degeneracy 3, then $\lambda _{1}=\lambda _{2}=\lambda
_{3}<\lambda _{4}$. This Schatten decomposition is not unique unless every
eigenvalue is non-degenerated. Then the entropy (von Neumann entropy\cite{OP}%
) $S\left( \rho \right) $ of a state $\rho $ is defined by

\begin{equation}
\renewcommand{\theequation}{2.1}
S\left( \rho \right) =-tr\rho \log \rho .
\end{equation}

The quantum mutual entropy was introduced on the basis of the above von
Neumann entropy for purely quantum communication processes. The mutual
entropy depends on an input state $\rho $ and a channel $\Lambda ^{*}$, so
it is denoted by $I\left( \rho ;\Lambda ^{*}\right) $, which should satisfy
the following conditions:

(1) The quantum mutual entropy is well-matched to the von Neumann entropy.
Furthermore, if a channel is trivial, i.e., $\Lambda ^{*}=$ identity map,
then the mutual entropy equals to the von Neumann entropy: $I\left( \rho
;id\right) $ = $S\left( \rho \right) $.

(2) When the system is classical, the quantum mutual entropy reduces to
classical one.

(3) Shannon's fundamental inequality \\ 0 $\leq $ $I\left( \rho
;\Lambda ^{*}\right) \leq S\left( \rho \right) $ is held.

In order to define the quantum mutual entropy followed by the classical one
(see\cite{O11}for the details), we need the joint state (it is called
``compound state'' in the sequel) describing the correlation between an
input state $\rho $ and the output state $\Lambda ^{*}\rho $ and the quantum
relative entropy. A finite partition of the classical measurable space
corresponds to an orthogonal decomposition $\left\{ E_{k}\right\} $ of the
identity operator I of $\mathcal{H}$ in quantum case because the set of all
orthogonal projections is considered to make an event system for a quantum
system. It is known \cite{OP}that the following equality holds

\[
\sup \left\{ -\sum_{k}tr\rho E_{k}\log tr\rho E_{k};\left\{ E_{k}\right\}
\right\} =-tr\rho \log \rho ,
\]
and the supremum is attained when $\left\{ E_{k}\right\} $ is a Schatten
decomposition of $\rho .$ Therefore the Schatten decomposition is used to
define the compound state and the quantum mutual entropy.

The compound state $\theta _{E}$ (corresponding to joint state
(measure) in CS) of $\rho $ and $\Lambda ^{*}\rho $ was introduced
in
\cite{O2,O3}, which is given by

\begin{equation}
\renewcommand{\theequation}{2.2}
\theta _{E}=\sum_{k}\lambda _{k}E_{k}\otimes \Lambda ^{*}E_{k},
\end{equation}
where $E$ stands for a Schatten decomposition of $\rho ,$ so that the
compound state depends on how we decompose the state $\rho $ into basic
states (elementary events).

The relative entropy for two states $\rho $ and $\sigma $ is defined by
Umegaki and Lindblad, which is written as

\begin{eqnarray*}
&&S\left( \rho ,\sigma \right)\\
&&= \left\{
\begin{array}{ll}
tr\rho \left( \log \rho -\log \sigma \right) & \left( \mbox{when
}\overline{ ran\rho }\subset \overline{ran\sigma }\right) \\
\infty & \left( \mbox{otherwise}\right)
\end{array}
\right..\\
&&\mbox{\hspace{6cm}(2.3)}
\end{eqnarray*}

Then we can define the quantum mutual entropy by means of the compound state
and the relative entropy \cite{O2}, that is,

\begin{equation}
\renewcommand{\theequation}{2.4}
I\left( \rho ;\Lambda ^{*}\right) =\sup \left\{ S\left( \theta _{E},\rho
\otimes \Lambda ^{*}\rho \right) ;E=\left\{ E_{k}\right\} \right\} ,
\end{equation}
where the supremum is taken over all Schatten decompositions. Some
computations reduce it to the following form:

\begin{equation}
\renewcommand{\theequation}{2.5}
I\left( \rho ;\Lambda ^{*}\right) =\sup \left\{ \sum_{k}\lambda _{k}S\left(
\Lambda ^{*}E_{k},\Lambda ^{*}\rho \right) ;E=\left\{ E_{k}\right\} \right\}
\end{equation}
This mutual entropy satisfies all conditions (1)$\sim $(3) mentioned above%
\cite{O11}.

When the input system is classical, an input state $\rho $ is given by a
probability distribution or a probability measure, in either case, the
Schatten decomposition of $\rho $ is unique, namely, for the case of
probability distribution ; $\rho =\left\{ \lambda _{k}\right\} ,$

\begin{equation}
\renewcommand{\theequation}{2.6}
\rho =\sum_{k}\lambda _{k}\delta _{k},
\end{equation}
where $\delta _{k}$ is the delta measure, that is, $\delta _{k}\left(
j\right) =\delta _{k,j}=\{_{0(k\neq j)}^{1(k=j)},\forall j.$ Therefore for
any channel $\Lambda ^{*},$ the mutual entropy becomes

\begin{equation}
\renewcommand{\theequation}{2.7}
I\left( \rho ;\Lambda ^{*}\right) =\sum_{k}\lambda _{k}S\left( \Lambda
^{*}\delta _{k},\Lambda ^{*}\rho \right) ,
\end{equation}
which equals to the following usual expression of Shannon when it is
well-defined:

\begin{equation}
\renewcommand{\theequation}{2.8}
I\left( \rho ;\Lambda ^{*}\right) =S\left( \Lambda ^{*}\rho \right)
-\sum_{k}\lambda _{k}S\left( \Lambda ^{*}\delta _{k}\right) ,
\end{equation}
which has been taken as the definition of the mutual entropy for a
classical-quantum(-classical) channel \cite{BS,Hol,Ing,Lev}.

Note that the above definition of the mutual entropy (2.5) is
also written as

\begin{eqnarray*}
I\left( \rho ;\Lambda ^{*}\right)&=&  \sup \left\{ \sum_{k}\lambda
_{k}S\left( \Lambda ^{*}\rho _{k},\Lambda ^{*}\rho \right) \right. \\
&&\qquad\left. ;\rho =\sum_{k}\lambda _{k}\rho _{k}\in F_{o}\left(
\rho
\right) \right\} ,
\end{eqnarray*}
where $F_{o}\left( \rho \right) $ is the set of all orthogonal finite
decompositions of $\rho $ \cite{O11}$.$

More general mutual entoropy was defined in \cite{O4} based on
Araki's relative entoropy \cite{Ara}.

\section{Communication Processes}

The information communication process is mathematically set as follows: M
messages are sent to a receiver and the $k$th message $\omega ^{\left(
k\right) }$ occurs with the probability $\lambda _{k}$. Then the occurence
probability of each message in the sequence $\left( \omega ^{\left( 1\right)
},\omega ^{\left( 2\right) },\right.$ $\left.\cdot \cdot \cdot ,\omega ^{\left(
M\right) }\right) $of M messages is denoted by $\rho =\left\{ \lambda
_{k}\right\} ,$ which is a state in a classical system. If $\xi $ is a classical
coding, then $\xi \left( \omega \right) $ is a classical object such as an e
lectric
pulse. If $\xi $ is a quantum coding, then $\xi \left( \omega \right) $ is a
quantum object (state) such as a coherent state. Here we consider such a
quantum coding, that is, $\xi \left( \omega ^{\left( k\right) }\right) $ is
a quantum state, and we denote $\xi \left( \omega ^{\left( k\right) }\right)
$ by $\sigma _{k}.$ Thus the coded state for the sequence $\left( \omega
^{\left( 1\right) },\omega ^{\left( 2\right) },\cdot \cdot \cdot ,\omega
^{\left( M\right) }\right) $ is written as $\sigma =\sum_{k}\lambda
_{k}\sigma _{k}.$ This state is transmitted through a channel $\gamma $,
which is expressed by a completely positive mapping $\Gamma ^{*}$ from the
state space of $X$ to that of $\stackrel{\sim }{X}$ , hence the output coded
quantum state $\stackrel{\sim }{\sigma }$ is $\Gamma ^{*}\sigma .$ Since the
information transmission process can be understood as a process of state
(probability) change, when $\Omega $ and $\stackrel{\sim }{\Omega }$ are
classical and $X$ and $\stackrel{\sim }{X}$ are quantum, the process is
written as

\begin{equation}
\renewcommand{\theequation}{3.1}
P\left( \Omega \right) \stackrel{\Xi ^{*}}{\longrightarrow }\mathcal{S}
\left( \mathcal{H}\right) \stackrel{\Gamma ^{*}}{\longrightarrow }\mathcal{S(%
}\stackrel{\sim }{\mathcal{H}})\stackrel{\stackrel{\sim }{\Xi }^{*}}{%
\longrightarrow }P(\stackrel{\sim }{\Omega }),
\end{equation}
where $\Xi ^{*}$ $($resp.$\stackrel{\sim }{\Xi }^{*})$ is the channel
corresponding to the coding $\xi $ (resp. decoding $\stackrel{\sim }{\xi }$
).

We have to be care to study the objects in the above transmission process
(3.1). For instance, if we want to know the information capacity of a
quantum channel $\gamma (=\Gamma ^{*}),$ then we have to take $X$ so as to
describe a quantum system like a Hilbert space and we need to start the
study from a quantum state in quantum space $X\ $not from a classical state
associated to a message. If we like to know the capacity of the whole
process including a coding and a decoding, which means the capacity of a
channel $\stackrel{\sim }{\xi }\circ \gamma \circ \xi (=\stackrel{\sim }{\Xi
}^{*}\circ \ \Gamma ^{*}\circ \Xi ^{*})$, then we have to start from a
classical state$.$

\section{Channel Capacity}

We discuss two types of channel capacity in communication processes, namely,
the capacity of a quantum channel $\Gamma ^{*}$ and that of a classical
(classical-quantum-classical) channel $\stackrel{\sim }{\Xi }^{*}\circ \
\Gamma ^{*}\circ \Xi ^{*}.$

(1) {\it Capacity of quantum channel:} The capacity of a quantum
channel is the ability of information transmission of a quantum
channel itself, so that it does not depend on how to code a message
being treated as classical object and we have to start from an
arbitrary quantum state and find the supremum of the quantum
mutual entropy. One often makes a mistake in this point. For
example, one starts from the coding of a message and compute the
supremum of the mutual entropy and he says that the supremum is
the capacity of a quantum channel, which is not correct. Even when
his coding is a quantum coding and he sends the coded message to a
receiver through a quantum channel, if he starts from a classical
state, then his capacity is not the capacity of the quantum channel
itself. In his case, usual Shannon's theory is applied because he can
easily compute the conditional probability by a usual (classical) way.
His supremum is the capacity of a classical-quantum-classical
channel, and it is in the second category discussed below.

The capacity of a quantum channel $\Gamma ^{*}$ is defined as follows: Let $%
\mathcal{S}_{0}(\subset $ $\mathcal{S(H))}$ be the set of all states
prepared for expression of information. Then the capacity of the channel $%
\Gamma ^{*}$ with respect to $\mathcal{S}_{0}$ is defined by

\begin{equation}
\renewcommand{\theequation}{4.1}
C^{\mathcal{S}_{0}}\left( \Gamma ^{*}\right) =\sup \{I\left( \rho ;\Gamma
^{*}\right) ;\rho \in \mathcal{S}_{0}\}.
\end{equation}
Here $I\left( \rho ;\Gamma ^{*}\right) $ is the mutual entropy given in
(2.4) or (2.5) with $\Lambda ^{*}=\Gamma ^{*}.$ When $\mathcal{S}_{0}=%
\mathcal{S(H)}$ , $C^{\mathcal{S}(\mathcal{H)}}\left( \Gamma ^{*}\right) $
is denoted by $C\left( \Gamma ^{*}\right) $ for simplicity.

In \cite{OPW1,Mur-O1}, we also considered the pseudo-quant-um capacity
$C_{p}\left( \Gamma ^{*}\right)$ defined by (4.1) with the pseudo-mutual
entropy $I_{p}\left( \rho ;\Gamma ^{*}\right)$ where the supremum is taken
over all finite decompositions instead of all orthogonal pure
decompositions:
\begin{eqnarray*}
&&I_{p}\left( \rho ;\Gamma ^{*}\right)  \nonumber \\
&=&\sup \left\{ \sum_{k}\lambda _{k}S\left( \Gamma ^{*}\rho _{k},\Gamma
^{*}\rho \right) ;\rho =\sum_{k}\lambda _{k}\rho _{k},\right.  \nonumber \\
&&\qquad\left. \mbox{ finite decomposition}\right\} .
\quad\quad\quad\quad \mbox{(4.2)}
\end{eqnarray*}

However the pseudo-mutual entropy is not well-matched to the conditions
explained in Sec.2, and it is difficult to be computed numerically. The
relation between $C\left( \Gamma ^{*}\right) $ and $C_{p}\left( \Gamma
^{*}\right) $ was discussed in\cite{OPW1}. From the monotonicity of the
mutual entropy\cite{OP}, we have

\[
0\leq C^{\mathcal{S}_{0}}\left( \Gamma ^{*}\right) \leq C_{p}^{\mathcal{S}
_{0}}\left( \Gamma ^{*}\right) \leq \sup \left\{ S(\rho );\rho \in \mathcal{S%
}_{0}\right\} .
\]

(2) {\it Capacity of classical-quantum-classical channel:} The
capacity of C-Q-C channel $\stackrel{\sim }{\Xi }^{*}\circ \ \Gamma
^{*}\circ \Xi ^{*} $ is the capacity of the information transmission
process starting from the coding of messages, therefore it can be
considered as the capacity including a coding (and a decoding). As is
discussed in Sec.3, an input state $\rho $ is the probability
distribution $\left\{ \lambda _{k}\right\} $ of messages, and its
Schatten decomposition is unique, so the mutual entropy is written
by (2.7):

\begin{eqnarray*}
&&I\left( \rho ;\stackrel{\sim }{\Xi }^{*}\circ \ \Gamma ^{*}\circ \Xi
^{*}\right)   \\
&=&\sum_{k}\lambda _{k}S\left( \stackrel{\sim }{\Xi }^{*}\circ \
\Gamma ^{*}\circ \Xi ^{*}\delta _{k},\stackrel{\sim }{\Xi }^{*}\circ \
\Gamma ^{*}\circ \Xi ^{*}\rho \right)\\
&&
\mbox{\hspace{6cm}(4.3)}
\end{eqnarray*}
If the coding $\Xi ^{*}$ is a quantum coding, then $\Xi ^{*}\delta _{k}$ is
expressed by a quantum state. Let denote the coded quantum state by $\sigma
_{k}$ and put $\sigma =\Xi ^{*}\rho =\sum_{k}\lambda _{k}\sigma _{k}.$ Then
the above mutual entropy is written as

\begin{eqnarray*}
&&I\left( \rho ;\stackrel{\sim }{\Xi }^{*}\circ \ \Gamma ^{*}\circ \Xi
^{*}\right) \\
&=&\sum_{k}\lambda _{k}S\left( \stackrel{\sim }{\Xi
}^{*}\circ \
\Gamma ^{*}\sigma _{k},\stackrel{\sim }{\Xi }^{*}\circ \ \Gamma ^{*}\sigma
\right) . \quad\quad\mbox{(4.4)}
\end{eqnarray*}
This is the expression of the mutual entropy of the whole information
transmission process starting from a coding of classical messages. Hence the
capacity of C-Q-C channel is

\begin{eqnarray*}
&&C^{P_{0}}\left( \stackrel{\sim }{\Xi }^{*}\circ \ \Gamma ^{*}\circ \Xi
^{*}\right) \\
&=&\sup \{I\left( \rho ;\stackrel{\sim }{\Xi }^{*}\circ \ \Gamma
^{*}\circ \Xi ^{*}\right) ;\rho \in P_{0}\}, \quad\quad\mbox{(4.5)}
\end{eqnarray*}
where $P_{0}(\subset P(\Omega ))$ is the set of all probability
distributions prepared for input (a-priori) states (distributions or
probability measures). Moreover the capacity for coding free is found by
taking the supremum of the mutual entropy over all probability distributions
and all codings $\Xi ^{*}$:

\begin{eqnarray*}
&&C_{c}^{P_{0}}\left( \stackrel{\sim }{\Xi }^{*}\circ \ \Gamma ^{*}\right) \\
&=&\sup \{I\left( \rho ;\stackrel{\sim }{\Xi }^{*}\circ \ \Gamma ^{*}\circ \Xi
^{*}\right) ;\rho \in P_{0},\Xi ^{*}\}. \quad\,\mbox{(4.6)}
\end{eqnarray*}
The last capacity is for both coding and decoding free and it is given by

\begin{eqnarray*}
&&C_{cd}^{P_{0}}\left( \ \Gamma ^{*}\right) \\
&=&\sup \{I\left( \rho ;\stackrel{%
\sim }{\Xi }^{*}\circ \ \Gamma ^{*}\circ \Xi ^{*}\right) ;\rho \in P_{0},\Xi
^{*},\stackrel{\sim }{\Xi }^{*}\}. \mbox{(4.7)}
\end{eqnarray*}
These capacities $C_{c}^{P_{0}},$ $C_{cd}^{P_{0}}$ do not measure the
ability of the quantum channel $\Gamma ^{*}$ itself, but measure the ability
of $\Gamma ^{*}$ through the coding and decoding.

Remark that $\sum_{k}\lambda _{k}S(\Gamma ^{*}\sigma _{k})$ is finite, then
(4.4) becomes

\begin{eqnarray*}
&&I\left( \rho ;\stackrel{\sim }{\Xi }^{*}\circ \ \Gamma ^{*}\circ \Xi
^{*}\right) \\
&=&S(\stackrel{\sim }{\Xi }^{*}\circ \Gamma ^{*}\sigma
)-\sum_{k}\lambda _{k}S(\stackrel{\sim }{\Xi }^{*}\circ \Gamma ^{*}\sigma
_{k}). \quad\quad\mbox{(4.8)}
\end{eqnarray*}
Further, if $\rho $ is a probability measure having a density function $%
f(\lambda )$ and each $\lambda $ corresponds to a quantum coded state $%
\sigma (\lambda ),$ then $\sigma =\int f(\lambda )$ $\sigma (\lambda
)d\lambda $ and

\begin{eqnarray*}
\renewcommand{\theequation}{4.9}
&&I\left( \rho ;\stackrel{\sim }{\Xi }^{*}\circ \ \Gamma ^{*}\circ \Xi
^{*}\right)  \\
&=&S(\stackrel{\sim }{\Xi }^{*}\circ \Gamma ^{*}\sigma )-\int
f(\lambda )S(\stackrel{\sim }{\Xi }^{*}\circ \Gamma ^{*}\sigma
(\lambda ))d\lambda . \\
&&\mbox{\hspace{6cm}(4.9)}
\end{eqnarray*}
This is bounded by

\[
S(\Gamma ^{*}\sigma )-\int f(\lambda )S(\Gamma ^{*}\sigma (\lambda
))d\lambda ,
\]
which is called the Holevo bound and is computed in several ocassions\cite
{YO,OPW1}.

The above three capacities $C^{P_{0}},$ $C_{c}^{P_{0}},$ $C_{cd}^{P_{0}}$
satisfy the following inequalities
\begin{eqnarray*}
0 &\leq &C^{P_{0}}\left( \stackrel{\sim }{\Xi }^{*}\circ \ \Gamma ^{*}\circ
\Xi ^{*}\right) \leq C_{c}^{P_{0}}\left( \stackrel{\sim }{\Xi }^{*}\circ \
\Gamma ^{*}\right) \\
&\leq &C_{cd}^{P_{0}}\left( \ \Gamma ^{*}\right) \leq \sup \left\{ S(\rho
);\rho \in P_{o}\right\}
\end{eqnarray*}
where $S(\rho )$ is not the von Neumann entropy but the Shannon entropy: -$%
\sum \lambda _{k}\log \lambda _{k}.$

The capacities (4.1), (4.5),(4.6) and (4.7) are generally different. Some
misuses occur due to forgetting which channel is considered. That is, we
have to make clear what kind of the ability (capacity) is considered, the
capacity of a quantum channel itself or that of a
classical-quantum(-classical ) channel. The computation of the capacity of a
quantum channel was carried in several models in \cite{OPW1}

\section{Compound States and Entanglements}

Recently the quantum entangled state has been mathematically studied \cite
{BBPSSW,Maj,Sch1}, in which the entangled state is defined by a state not
written as a form $\sum_{k}\lambda _{k}\rho _{k}\otimes \sigma _{k}$ with
any states $\rho _{k}$ and $\sigma _{k}.$ A state written as above is called
a separable state, so that an entangled state is a state not belonged to the
set of all separable states. However it is obvious that there exist several
correlated states written as separable forms. Such correlated states have
been discussed in several contexts in quantum probability such as quantum
filtering \cite{B2}, quantum compound state \cite{O2}, quantum Markov state
\cite{A} and quantum lifting \cite{AO}. In \cite{BO}, we showed a
mathematical construction of quantum entangled states and gave a finer
classification of quantum sates.

For the (separable) Hilbert space $\mathcal{K}$ of a quantum system, let $%
\mathcal{A\equiv }$ $B\left( \mathcal{K}\right) $ be the set of all linear
bounded operators {on }$\mathcal{K}$. A normal state $\varphi $ on $\mathcal{%
\ \ A}$ can be expressed as $\varphi \left( A\right) =tr_{\mathcal{G}}\kappa
^{\dagger }A\kappa ,$ $A\in \mathcal{A}$, where $\mathcal{G}$ is another
separable Hilbert space, $\kappa $ is a linear Hilbert-Schmidt operator from
$\mathcal{G}$ to $\mathcal{K}$ and $\kappa ^{\dagger }$ is the adjoint
operator of $\kappa $ from $\mathcal{K}$ to $\mathcal{G}$ such that $\sigma
=\kappa \kappa ^{\dagger }$ is the (unique) density operator $\sigma \in
\mathcal{A}$ of the state $\varphi :\varphi \left( A\right) =trA\sigma $, $%
A\in \mathcal{A}$. This $\kappa $ is called the amplitude operator, and it
is called just the amplitude if $\mathcal{G}$ is one dimensional space
${\bf C}$, corresponding to the pure state $\varphi \left( A\right) =\kappa
^{\dagger }A\kappa $ for a $\kappa \in \mathcal{K}$ with $\kappa ^{\dagger
}\kappa =\Vert \kappa \Vert ^{2}=1$. In general, $\mathcal{G}$ is not one
dimensional, the dimensionality $\dim \mathcal{G}$ must be not less than $%
\mathrm{rank}\sigma $, the dimensionality of the range $\sigma \mathcal{K}$
of the density operator $\sigma .$

Since $\mathcal{G}$ is separable, $\mathcal{G}$ is realized as a subspace of
$l^{2}(\mathbf{N})$ of complex sequences $(i.e.,\zeta ^{\bullet }=\left(
\zeta ^{n}\right) ,$ $\zeta ^{n}\in {\bf C}$, $n\in \mathbf{N}$ with $\sum
\left| \zeta ^{n}\right| ^{2}<+\infty )$, so that any vector $\zeta
^{\bullet }=(\zeta ^{n}$) represents a vector $\zeta =\sum \zeta
^{n}|n\rangle $ in the standard basis $\left\{ |n\rangle \right\} \in
\mathcal{G}$ of $l^{2}(\mathbf{N})$ .

Given the amplitude operator $\kappa $, one can define not only the states $%
\sigma \equiv \kappa \kappa ^{\dagger }$ and $\rho \equiv $ $\kappa
^{\dagger }\kappa $ on the algebras $\mathcal{A}\left( =B\left( \mathcal{K}%
\right) \right) $ and $\mathcal{B}\left( =B\left( \mathcal{G}\right) \right)
$ but also an entanglement state $\Theta $ on the algebra $\mathcal{B}%
\otimes \mathcal{A}$ of all bounded operators on the tensor product Hilbert
space $\mathcal{G}\otimes \mathcal{K}$ by

\[
\Theta \left( B\otimes A\right) =tr_{\mathcal{G}}B\kappa ^{\dagger }A\kappa
=tr_{\mathcal{K}}A\kappa B\kappa ^{\dagger },
\]
for any $B\in \mathcal{B}$. This state is pure as it is the case of $%
\mathcal{F}={\bf C}$ in the theorem below, and it satisfies the marginal
conditions: For any $B\in \mathcal{B},A\in \mathcal{A}$,

\[
\Theta \left( B\otimes I\right) =tr_{\mathcal{G}}B\rho ,\quad \Theta \left(
I\otimes A\right) =tr_{\mathcal{K}}A\sigma .\quad
\]

\bigskip

\noindent
{\bf Theorem 5.1.} {\sl
Let $\Theta :\mathcal{B}\otimes \mathcal{A}\rightarrow {\bf C}$ be a state
\[
\Theta \left( B\otimes A\right) =tr_{\mathcal{E}}\psi ^{\dagger }\left(
B\otimes A\right) \psi ,
\]
defined by an amplitude operator $\psi $ on a separable Hilbert space $%
\mathcal{E}$ into the tensor product Hilbert space $\mathcal{G}\otimes
\mathcal{K}$ ; $\psi :\mathcal{E}\rightarrow \mathcal{G}\otimes \mathcal{K}$
with $tr_{\mathcal{F}}\psi ^{\dagger }\psi =1$. Then there exist a Hilbert
space $\mathcal{F}$ and an amplitude operator $\kappa :\mathcal{G}%
\rightarrow \mathcal{F}\otimes \mathcal{K}$ with
\begin{equation}
\renewcommand{\theequation}{5.1}
\kappa ^{\dagger }\left( I\otimes \mathcal{A}\right) \kappa \subset \mathcal{%
\ B},\;tr_{\mathcal{F}}\kappa \mathcal{B}\kappa ^{\dagger }\subset \mathcal{A%
}
\end{equation}
such that the state $\Theta $ can be achieved by an entanglement
\begin{eqnarray*}
\Theta \left( B\otimes A\right) &=&tr_{\mathcal{G}}B\kappa ^{\dagger }\left(
I\otimes A\right) \kappa \\
& =&tr_{\mathcal{F}\otimes \mathcal{K}}\left( I\otimes
A\right) \kappa B\kappa ^{\dagger } \mbox{\hspace{1.3cm}{\rm (5.2)}}
\end{eqnarray*}
The entangling operator $\kappa $ is uniquely defined up to a unitary
transformation of the minimal space $\mathcal{F}$.}

\bigskip

Note that the entangled state (5.2) is written as
\begin{equation}
\renewcommand{\theequation}{5.3}
\Theta \left( B\otimes A\right) =tr_{\mathcal{G}}B\phi \left( A\right) =tr_{%
\mathcal{K}}A\phi _{*}\left( B\right) ,
\end{equation}
where $\phi \left( A\right) =\kappa ^{\dagger }\left( I\otimes A\right)
\kappa $ is in the predual space $\mathcal{B}_{*}\subset \mathcal{B}$ of all
trace-class operators in $\mathcal{G}$, and $\phi _{*}\left( B\right) =tr_{%
\mathcal{F}}\kappa B\kappa ^{\dagger }$ is in $\mathcal{A}_{*}\subset
\mathcal{A}$. The map $\phi $ is the Steinspring form of the general
completely positive map $\mathcal{A}\rightarrow \mathcal{B}_{*}$, written in
the eigen-basis $\left\{ \left| n\right\rangle \right\} $ of $\mathcal{G}
\subseteq l^{2}\left( {\bf N}\right) $ of the density operator $\rho =\phi
\left( I\right) $ as
\begin{equation}
\renewcommand{\theequation}{5.4}
\phi \left( A\right) =\sum_{m,n}|m\rangle \kappa _{m}^{\dagger }\left(
I\otimes A\right) \kappa _{n}\langle n|,\quad A\in \mathcal{A}
\end{equation}
where $\kappa _{n}$ is the vector in $\mathcal{F}\otimes \mathcal{K}$ such
that $\kappa =\sum_{n}\kappa _{n}\langle n|$. The dual operation $\phi _{*}$
is the Kraus form of the general completely positive map $\mathcal{B}%
\rightarrow \mathcal{A}_{*}$, given in this basis as
\begin{equation}
\renewcommand{\theequation}{5.5}
\phi _{*}\left( B\right) =\sum_{n,m}\left\langle n\right| B\left|
m\right\rangle tr_{\mathcal{F}}\kappa _{n}\kappa _{m}^{\dagger },\quad B\in
\mathcal{B}.
\end{equation}
It corresponds to the general form of the density operator
\begin{equation}
\renewcommand{\theequation}{5.6}
\theta _{\phi }=\sum_{m,n}|n\rangle \langle m|\otimes tr_{\mathcal{F}}\kappa
_{n}\kappa _{m}^{\dagger }
\end{equation}
for the entangled state $\Theta $, characterized by the weak orthogonality
property
\begin{equation}
\renewcommand{\theequation}{5.7}
tr_{\mathcal{K}}\kappa _{n}\kappa _{m}^{\dagger }=p_{n}\delta
_{n}^{m}=\kappa _{m}^{\dagger }\kappa _{n}.
\end{equation}

\bigskip

\noindent
{\bf Definition 5.2.} {\sl
The dual map $\phi _{*}:\mathcal{B}\rightarrow \mathcal{A}_{*}$ to a
completely positive map $\phi :\mathcal{A}\rightarrow \mathcal{B}_{*}$,
normalized as $tr_{\mathcal{G}}\phi \left( I\right) =1$, is called the
quantum entanglement of the state $\rho =\phi \left( I\right) $ on $\mathcal{%
B}$ to the state $\sigma =\phi _{*}\left( I\right) $ on $\mathcal{A}$. The
entanglement by $\phi \left( A\right) =\sigma ^{1/2}A\sigma ^{1/2}$ of the
state $\rho =\sigma $ on the algebra $\mathcal{B}=\mathcal{A}$ given by the
standard entangling operator $\kappa =\sigma ^{1/2}$ is called standard.}

\bigskip

\section{d-Entanglements and Correspondences}

A compound state, playing the similar role as the joint input-output
probability measures in classical systems, was intorduced in \cite{O2} as
explained in Sec.2. It corresponds to a particular diagonal type
\[
\theta _{\phi }=\sum_{n}|n\rangle \langle n|\otimes \kappa _{n}\kappa
_{n}^{\dagger }
\]
of the entangling map (5.6) in the eigen-basis of the density operator
$%
\rho =\sum p_{n}|n\rangle \langle n|$, and is discussed in this section.
Therefore the entangled states, generalizing the compound state, also play
the role of the joint probability measures.

The diagonal entanglements are quantum correspondences of classical symbols
to quantum, in general not orthogonal and pure, states. The general
entangled states $\Theta $ are described by the density operators $\theta
_{\phi }$ of the form (5.6) which is not necessarily diagonal in the
eigen-representation of the density operator $\rho =\sum_{n}p_{n}|n\rangle
\langle n|$. Such nondiagonal entangled states were called in \cite{O4} the
quasicompound (q-compound) states, so we can call also the nondiagonal
entanglement the quantum quasi-correspondence (q-correspondece) in contrast
to the d-correspondences, described by the diagonal entanglements, giving
rise to the d-compound states.

Let us consider a finite or infinite input system indexed by the natural
numbers $n\in \mathbf{N}$. The associated space $\mathcal{G}\subseteq
l^{2}\left( \mathbf{N}\right) $ is the Hilbert space of the input system
described by a quantum projection-valued measure $n\mapsto |n\rangle \langle
n|$ on $\mathbf{N}$ giving an orthogonal partition of unity $I=\sum
|n\rangle \langle n|$ $\in \mathcal{B}$ of the finite or infinite
dimensional input Hilbert space $\mathcal{G}$. Each input pure state,
identified with the one-dimensional density operator $|n\rangle \langle
n|\in \mathcal{B}$ corresponding to the elementary symbol $n\in \mathbf{N}$,
defines the elementary output state $\omega _{n}$ on $\mathcal{A}$. If the
elementary states $\omega _{n}$ are pure, they are described by pure output
amplitudes $\upsilon _{n}\in \mathcal{K}$ satisfying $\upsilon _{n}^{\dagger
}\upsilon _{n}=1=tr\omega _{n}$, where $\omega _{n}=$ $\upsilon _{n}\upsilon
_{n}^{\dagger }$ are the corresponding output one-dimensional density
operators. If these amplitudes are non-orthogonal $\upsilon _{m}^{\dagger
}\upsilon _{n}\neq \delta _{n}^{m}$, they cannot be identified with the
input amplitudes $|n\rangle $.

The elementally joint input-output states are given by the density operators
$|n\rangle \langle n|\otimes \omega _{n}$ in $\mathcal{G}\otimes \mathcal{K}$%
, and their mixtures
\begin{equation}
\renewcommand{\theequation}{6.1}
\theta =\sum_{n}|n\rangle \langle n|\otimes \sigma _{n},\quad \sigma
_{n}=p_{n}\upsilon _{n}\upsilon _{n}^{\dagger }
\end{equation}
define the compound states on $\mathcal{B}\otimes \mathcal{A}$, giving the
quantum correspondences $n\mapsto |n\rangle \langle n|$ with the
probabilities $p_{n}$. Here we note that the quantum correspondence is
described by a classical-quantum channel, and the general d-compound state
for a quantum-quantum channel in quantum communication can be obtained in
this way due to the orthogonality of the decomposition (6.1),
corresponding to the orthogonality of the Schatten decomposition $\rho
=\sum_{n}p_{n}|n\rangle \langle n|$ of $\rho =tr_{\mathcal{K}}\theta $.

The comparison of the general compound state (5.6) with (6.1)
suggests that the quantum correspondences are described as the
diagonal entanglements
\begin{equation}
\renewcommand{\theequation}{6.2}
\phi _{*}\left( B\right) =\sum_{n}p_{n}\langle n|B|n\rangle \upsilon
_{n}\upsilon _{n}^{\dagger }
\end{equation}
which are dual to the orthogonal decompositions
\begin{equation}
\renewcommand{\theequation}{6.3}
\phi \left( A\right) =\sum_{n}p_{n}|n\rangle \upsilon _{n}^{\dagger
}A\upsilon _{n}\langle n|.
\end{equation}
\newline
These are the entanglements with the stronger orthogonality
\begin{equation}
\renewcommand{\theequation}{6.4}
tr_{\mathcal{F}}\kappa _{n}\kappa _{m}^{\dagger }=p_{n}\omega _{n}\delta
_{n}^{m},
\end{equation}
for the amplitudes $\kappa _{n}\in \mathcal{F}\otimes \mathcal{K}$ of the
decomposition $\kappa =\sum_{n}\kappa _{n}\langle n|$ in comparison with the
weak orthogonality of $\kappa _{n}$ in (5.7). The orthogonality (6.4)
can be achieved in the following manner: Take $\kappa _{n}=|n\rangle
\otimes \psi _{n}$ with $\psi _{n}=p_{n}^{1/2}\upsilon _{n}$ so that
\[
\kappa _{m}^{\dagger }\left( I\otimes A\right) \kappa _{n}=\left\langle
m\mid n\right\rangle \psi _{m}^{\dagger }A\psi _{n}=p_{n}\upsilon
_{n}^{\dagger }A\upsilon _{n}\delta _{n}^{m}
\]
for any $A\in \mathcal{A}$. Then, we have the following theorem.

\bigskip

\noindent
{\bf Theorem 6.1.} {\sl
Let $\mathcal{F=\oplus }_{n}\mathcal{F}_{n}$ and let $\psi _{n}$ be the
operators, defining a compound state of the diagonal form
\begin{equation}
\renewcommand{\theequation}{6.5}
\Theta \left( B\otimes A\right) =\sum_{n}\langle n|B|n\rangle tr_{\mathcal{F}%
_{n}}\psi _{n}^{\dagger }A\psi _{n}
\end{equation}
Then it corresponds to the entanglement by the orthogonal decomposition
\begin{equation}
\renewcommand{\theequation}{6.6}
\phi \left( A\right) =\sum_{n}|n\rangle \kappa _{n}^{\dagger }\left(
I\otimes A\right) \kappa _{n}\langle n|,
\end{equation}
mapping from the algebra $\mathcal{A}$ into a diagonal subalgebra of $%
\mathcal{B}$.}

\bigskip

Thus the entanglement (5.5) corresponding to (6.5) is given by
the dual to (6.6) diagonal map
\begin{equation}
\renewcommand{\theequation}{6.7}
\phi _{*}\left( B\right) =\sum_{n}\langle n|B|n\rangle \psi _{n}\psi
_{n}^{\dagger }
\end{equation}
with the density operators $\sigma _{n}=\psi _{n}\psi _{n}^{\dagger }$
normalized to the probabilities $p_{n}=tr_{\mathcal{K}}\psi _{n}\psi
_{n}^{\dagger }$.

\bigskip

\noindent
{\bf Definition 6.2.} {\sl
The positive diagonal map
\begin{equation}
\renewcommand{\theequation}{6.8}
\phi _{*}\left( B\right) =\sum_{n}\langle n|B|n\rangle \sigma _{n}
\end{equation}
into the subspace of trace-class operation $\mathcal{K}$ with $tr_{\mathcal{G%
}}\phi _{*}\left( I\right) =1$, is called quantum d-entanglement with the
input probabilities $p_{n}=tr_{\mathcal{K}}\sigma _{n}$ and the output
states $\omega _{n}=p_{n}^{-1}\sigma _{n}$, and the corresponding compound
state (2.2) is called d-compound state. The d-entanglement is called
c-entanglement and compound state is called c-compound if all density
operators $\sigma _{n}$ commute: $\sigma _{m}\sigma _{n}=\sigma _{n}\sigma
_{m}$ for all $m$ and $n$. }

\bigskip

Note that due to the commutativity of the operators $B\otimes I$ with $%
I\otimes A$ on $\mathcal{G}\otimes \mathcal{K}$, one can treat the
correspondences as the nondemolition measurements in $\mathcal{B}$ with
respect to $\mathcal{A}$. So, the compound state is the state prepared for
such measurements on the input $\mathcal{G}$. It coincides with the mixture
of the states, corresponding to those after the measurement without reading
the sent message. The set of all d-entanglements corresponding to a given
Schatten decomposition of the input state $\rho $ on ${\rm A}$ is
obviously convex with the extreme points given by the pure elementary output
states $\omega _{n}$ on $\mathcal{A}$, corresponding to a not necessarily
orthogonal decompositions $\sigma =\sum_{n}\sigma _{n}$ into one-dimensional
density operators $\sigma _{n}=p_{n}\omega _{n}.$

The orthogonal Schatten decompositions $\sigma =\sum_{n}p_{n}\omega _{n}$
correspond to the extreme points of c-entanglements which also form a convex
set with mixed commuting $\omega _{n}$ for a given Schatten decomposition of
$\sigma $. The orthogonal c-entanglements were used in \cite{AO} to
construct a particular type of Accardi's transition expectations \cite{A}
and to define the entropy in a quantum dynamical system via such transition
expectations\cite{BO}.

Thus we classified the entangled states into three categories, namely,
q-entangled state, d-entangled state and c-entangled state, and their
rigorous expressions were given.

\section{Quantum Mutual Entropy via Entanglements}

Let us consider the entangled mutual entropy by means of the above three
types compound states. We denote the quantum mutual entropy of the compound
state $\Theta $ achieved by an entanglement $\phi _{*}:$ $\mathcal{B}%
\rightarrow \mathcal{A}_{*}$ with the marginals
\begin{equation}
\renewcommand{\theequation}{7.1}
\Theta \left( B\otimes I\right) =tr_{\mathcal{G}}B\rho ,\;\Theta \left(
I\otimes A\right) =tr_{\mathcal{K}}A\sigma
\end{equation}
by $I_{\phi }\left( \rho ,\sigma \right) $ or $I_{\phi }\left( \mathcal{A},%
\mathcal{B}\right) $ and it is given as
\begin{equation}
\renewcommand{\theequation}{7.2}
I_{\phi }\left( \mathcal{\rho },\mathcal{\sigma }\right) =tr\theta _{\phi
}\left( \log \theta _{\phi }-\log \left( \rho \otimes \sigma \right) \right)
.
\end{equation}
Besides this quantity describes an information gain in a quantum system $%
\left( \mathcal{A},\sigma \right) $ via an entanglement $\phi _{*}$ with
another system ($\mathcal{B},\rho ),$ it is naturally treated as a measure
of the strength of an entanglement, having zero the value only for
completely disentangled states (7.1), corresponding to $\theta _{\phi
}=\rho \otimes \sigma $.

\bigskip

\noindent
{\bf Definition 7.1.} {\sl
The maximal quantum mutual entropy for a fixed state $\sigma $
\begin{equation}
\renewcommand{\theequation}{7.3}
H_{\sigma }\left( \mathcal{A}\right) =\sup \{I_{\phi }\left( \mathcal{A},%
\mathcal{B}\right) ;\phi _{*}\left( I\right) =\sigma \}
\end{equation}
is called q-entropy of the state $\sigma $. The differences
\begin{eqnarray*}
H_{\phi }\left( \mathcal{B}|\mathcal{A}\right)  &=&H_{\sigma }\left(
\mathcal{A}\right) -I_{\phi }\left( \mathcal{A},\mathcal{B}\right) , \\
D_{\phi }\left( \mathcal{B}|\mathcal{A}\right)  &=&S\left( \mathcal{\sigma }%
\right) -I_{\phi }\left( \mathcal{A},\mathcal{B}\right)
\end{eqnarray*}
are respectively called the q-conditional entropy on $\mathcal{B}$ with
respect to $\mathcal{A}$ and the degree of disentanglement for the compound
state $\phi $.}

\bigskip

$H_{\phi }\left( \mathcal{B}|\mathcal{A}\right) $ is obviously positive,
however\\ $D_{\phi }\left( \mathcal{B}|\mathcal{A}\right) $ has the positive
maximal value $S\left( \mathcal{\sigma }\right) =$ $\sup \left\{ D_{\phi
}\left( \mathcal{B}|\mathcal{A}\right) ; \phi _{*}\left( I\right)
=\sigma
\right\} $ and can achieve also a negative value
\begin{equation}
\renewcommand{\theequation}{7.4}
\inf \left\{ D_{\phi }\left( \mathcal{B}|\mathcal{A}\right) ;\phi _{*}\left(
I\right) =\sigma \right\} =S\left( \mathcal{\sigma }\right) -H_{\sigma
}\left( \mathcal{A}\right)
\end{equation}
for the entangled states \cite{BO}.

\bigskip

\noindent
{\bf Theorem 7.2. }{\sl
Let $\mathcal{A}$ be a discrete decomposable algebra $\oplus B\left(
\mathcal{K}_{i}\right) $ with a normal state $\sigma =\oplus \sigma _{i}$ ,
and $\mathcal{C}\subseteq \mathcal{A}$ be its center with probability
distribution $\mu =\oplus \mu _{i}$ induced by $\sigma .$ Then the q-entropy
is given by
\begin{equation}
\renewcommand{\theequation}{7.5}
H_{\sigma }\left( \mathcal{A}\right) =\sum_{i}\left( \mu _{i}\ln \mu
_{i}-2tr_{\mathcal{K}_{i}}\sigma _{i}\ln \sigma _{i}\right) ,
\end{equation}
It is positive, $H_{\sigma }\left( \mathcal{A}\right) \in [0,\infty ]$, and
if $\mathcal{A}$ is finite dimensional, it is bounded, $H_{\sigma }\left(
\mathcal{A}\right) \leq \dim \mathcal{A}$.}

\bigskip

Let us consider $\mathcal{G}$ as a Hilbert space describing a quantum input
system and $\mathcal{K}$ as its output Hilbert space. A quantum channel $%
\Lambda ^{*}$ sending each input state defined on $\mathcal{G}$ to an output
state defined on $\mathcal{K}.$ A deterministic quantum channel is given by
a linear isometry $\Upsilon $ $\mathrm{:}\mathcal{G}\rightarrow
$$\mathcal{K}$ with $\Upsilon ^{\dagger }\Upsilon =I_{0}$ ($I_{0}$
is the identify operator in $\mathcal{G}$) such that each input state
vector $\eta
\in \mathcal{G}$, $\left\| \eta \right\| =1$ is transmitted into an output
state vector $\Upsilon \eta \in \mathcal{K}$, $\left\| \Upsilon \eta
\right\| =1$. The mixtures $\rho =\sum_{n}p_{n}\omega _{n}$ of the pure
input states $\omega _{n}=\eta _{n}\eta _{n}^{\dagger }$ are sent into the
mixtures $\sigma =\sum_{n}p_{n}\sigma _{n}$ with pure states $\sigma
_{n}=\Upsilon \omega _{n}\Upsilon ^{\dagger }$. A noisy quantum channel
sends pure input states $\omega $ into mixed ones $\sigma =\Lambda
^{*}\omega $ given by the dual of the following completely positive map $%
\Lambda $%
\begin{equation}
\renewcommand{\theequation}{7.6}
\Lambda \left( A\right) =\Upsilon ^{\dagger }\left( I_{1}\otimes A\right)
\Upsilon ,\mathrm{\quad }A\in \mathrm{\mathcal{A}}
\end{equation}
where $\Upsilon $ is a linear isometry from $\mathcal{G}$ to $\mathcal{F}
_{1}\otimes \mathcal{K}$, $\Upsilon ^{\dagger }\left(
I_{1}\otimes I\right) \Upsilon =I_{0}$, and $I_{1}$ is the identity operator
in a separable Hilbert space $\mathcal{F}_{1}$ representing the quantum
noise. Each input mixed state $\rho $ $\in B\left( \mathcal{G}\right) $ is
transmitted into the output state $\sigma =\Lambda ^{*}\rho $ on $\mathcal{A}%
\subseteq B\left( \mathcal{K}\right) $, which is given by the density
operator
\begin{equation}
\renewcommand{\theequation}{7.7}
\sigma =tr_{\mathcal{F}_{1}}\Upsilon \rho \Upsilon ^{\dagger }\equiv \Lambda
^{*}\rho \in \mathcal{A}_{*}.
\end{equation}

We apply the proceeding discussion of the entanglement to the above
situation containing a channel $\Lambda ^{*}.$ For a given Schatten
decomposition $\rho =\sum_{n}p_{n}|n\rangle \langle n|\ $and the state $%
\sigma \equiv \Lambda ^{*}\rho ,$we can construct three entangled states of
the preceeding section:

(1) q-entanglement $\phi _{*}^{q}$ and q-compound state $\theta _{\phi }^{q}$
are given as

\begin{eqnarray*}
\phi _{*}^{q}(B) &=&\sum_{n,m}\left\langle n\mid B\mid m\right\rangle tr_{%
\mathcal{F}}\kappa _{n}\kappa _{m}^{\dagger } \\
\theta _{\phi }^{q} &=&\sum_{m,n}|n\rangle \langle m|\otimes tr_{\mathcal{F}%
}\kappa _{n}\kappa _{m}^{\dagger }
\end{eqnarray*}
with the marginals $\rho =\sum_{n}p_{n}|n\rangle \langle n|,$ $\sigma \equiv
\Lambda ^{*}\rho =tr_{\mathcal{G}}\theta _{\phi }^{q}$ and $tr_{\mathcal{K}%
}\kappa _{n}\kappa _{m}^{\dagger }=p_{n}\omega _{n}\delta _{n}^{m}=\kappa
_{m}^{\dagger }\kappa _{n}$ for $\omega _{n}=\Lambda ^{*}|n\rangle \langle
n|.$ Let $\mathcal{E}_{q}$ be the convex set of all completely positive maps
$\phi ^{q}$ .

(2) d-entanglement $\phi _{*}^{d}$ and d-compound state $\theta _{\phi }^{d}$
are given as

\begin{eqnarray*}
\phi _{*}^{d}(B) &=&\sum_{n}\left\langle n\mid B\mid n\right\rangle tr_{%
\mathcal{F}}\kappa _{n}\kappa _{n}^{\dagger } \\
\theta _{\phi }^{d} &=&\sum_{n}|n\rangle \langle n|\otimes tr_{\mathcal{F}%
}\kappa _{n}\kappa _{n}^{\dagger }
\end{eqnarray*}

\noindent with the same marginal conditions as (1). Let $\mathcal{E}_{d}$ be
the convex set of all completely positive maps $\phi ^{d}.$

(3) c-entanglement $\phi _{*}^{c}$ and c-compound state $\theta _{\phi }^{c}$
are same as those of (2) with commuting $\left\{ \omega _{n}\right\} .$ Let $%
\mathcal{E}_{c}$ be the convex set of all completely positive maps $\phi
^{c} $ .

Now, let us consider the entangled mutual entropy and the capasity of
quantum channel by means of the above three types of compound states.

\bigskip

\noindent
{\bf Definition 7.3.} {\sl
The mutual entoropy $I_{q}\left( \rho ,\Lambda ^{*}\right) $ and
q-capacity $C_{q}\left( \Lambda ^{*}\right) $ for a quantum
channel $\Lambda ^{*}$are defined by the supremums
\begin{eqnarray*}
I_{q}\left( \rho ,\Lambda ^{*}\right)  &=&\sup \left\{ S(\theta _{\phi
}^{q},\rho \otimes \Lambda ^{*}\rho );\phi ^{q}\in \mathcal{E}_{q}\right\} ,
\\
\;C_{q}\left( \Lambda ^{*}\right)  &=&\sup \left\{ I_{q}\left( \rho ,\Lambda
^{*}\right) ;\rho \right\} .
\end{eqnarray*}
The d-mutual entropy, d-capacity and c-mutual entropy, c-capacity are
defined as above using $\theta _{\phi }^{d}$ and $\theta _{\phi }^{c}$,
respectively.}

\bigskip

Note that due to $\mathcal{E}_{c}\subseteq \mathcal{E}_{d}\subseteq
\mathcal{E}_{q}$, we have the inequalities
\begin{eqnarray*}
I_{q}\left( \rho ,\Lambda ^{*}\right)  &\geq &I_{d}\left( \rho ,\Lambda
^{*}\right) \geq I_{c}\left( \rho ,\Lambda ^{*}\right) ,\; \\
C_{q}\left( \Lambda ^{*}\right)  &\geq &C_{d}\left( \Lambda ^{*}\right) \geq %
C_{c}\left( \Lambda ^{*}\right)
\end{eqnarray*}
for a deterministic channel ($\Lambda ^{*}=id$), the two lower mutual
entropies coincide with the von Neumann entropy:
\[
I_{d}\left( \rho ,id\right) =-tr\rho \log \rho =I_{c}\left( \rho ,id\right) .
\]
The capacity for such a channel is finite if $\mathcal{A}$ has a finite
rank, $C_{d}\left( \Lambda ^{*}\right) \leq \dim \mathcal{K}$. On the other
hand, the q-mutual entropy can achieve the q-entropy
\[
I_{q}\left( \rho ,id\right) =-2tr\rho \log \rho
\]
and its capacity is bounded by the dimension of the algebra $\mathcal{A}$,
$C_{q}\left( \Lambda ^{*}\right) \leq \dim \mathcal{A}$ which doubles the
d-capacity dim$\mathcal{K}$ when $\mathcal{A}=B\left( \mathcal{K}\right)$.

\end{document}